\documentclass{jfm}
\usepackage[T1]{fontenc}
\usepackage[latin9]{inputenc}
\usepackage{soul}
\usepackage{xcolor}
\usepackage{graphicx}
\usepackage{units}
\usepackage{amssymb}
\usepackage{amsmath}
\usepackage[authoryear]{natbib}
\PassOptionsToPackage{normalem}{ulem}
\usepackage{ulem}

\makeatletter
\providecommand{\tabularnewline}{\\}
\providecolor{lyxadded}{rgb}{0,0,1}
\providecolor{lyxdeleted}{rgb}{1,0,0}
\def\vec#1{\boldsymbol{#1}}

\makeatother

\begin{document}

\title[Elementary model of internal electromagnetic pinch-type instability]
{Elementary model of internal\\
electromagnetic pinch-type instability}

\author[J. Priede]{J\={a}nis Priede}
\affiliation{
Applied Mathematics Research Centre,\\
Coventry University, UK}

\maketitle

\begin{abstract}
We analyse numerically a pinch-type instability in a semi-infinite
planar layer of inviscid conducting liquid bounded by solid walls
and carrying a uniform electric current. Our model is as simple as
possible but still captures the salient features of the instability
which otherwise may be obscured by the technical details of more comprehensive
numerical models and laboratory experiments. Firstly, we show the
instability in liquid metals, which are relatively poor conductors,
differs significantly from the astrophysically-relevant Tayler instability.
In liquid metals, the instability develops on the magnetic response
time scale, which depends on the conductivity and is much longer than
the Alfvén time scale, on which the Tayler instability develops in
well conducting fluids. Secondly, we show that this instability is
an edge effect caused by the curvature of the magnetic field, and
its growth rate is determined by the linear current density and independent
of the system size. Our results suggest that this instability may
affect future liquid metal batteries when their size reaches a few
meters.
\end{abstract}

\section{Introduction}

The electromagnetic force that results from the interaction of electric
current with its own magnetic field is usually rotational and, thus,
induces a flow when the conducting medium is fluid \citep{Boiarevich1989}.
However, some rotationally or translationally invariant current distributions
can produce a purely potential electromagnetic force which can be
balanced by the pressure gradient alone. Such quiescent equilibrium
states are not always stable and can collapse when subject to an arbitrary
small disturbance. The prominent example is the electromagnetic pinch,
which can disrupt or kink conductors carrying strong electric currents
\citep{Haines2000}. It usually affects conducting media with deformable
boundaries and has a growth rate determined by the balance of inertia
and the characteristic magnetic pressure. This results in the instability
that develops with the Alfvén wave speed regardless of the electrical
conductivity of the fluid \citep{Tayler1960}. Electromagnetic pinch-type
instability can develop also in the incompressible liquids bounded
by solid walls \citep{Michael1954}. In a perfectly conducting liquid,
this internal pinch instability also develops on the Alfvén time scale
\citep{Velikhov1959}. The previous two studies were extended by \citet{Tayler1961}
to general non-axisymmetric instability modes in a perfectly conducting
fluid bounded by solid cylindrical walls. Later it was suggested by
\citet{Vandakurov1972} and \citet{Tayler1973} that a similar instability
can affect the interiors of stars containing toroidal magnetic fields.
The strong stratification in stellar interiors makes this instability
nearly horizontal and, thus, significantly different from the pinch
instabilities with deformable boundaries \citep{Tayler1957}. In order
to stress this difference, \citet{Spruit1999} termed the astrophysical
variety of pinch instability the Tayler instability. This term was
later used by \citet{Ruediger2007} in a much broader sense to refer
to current-driven instabilities in homogeneous fluids including liquid
metals. The latter type of instability, which differs from the Tayler
instability not only by the absence of radial stratification but also
by its resistive nature, which will be discussed in this paper, was
presumably observed in the liquid-metal experiment by \citet{Seilmayer2012}.
The notion of Tayler instability is further broadened in the recent
study by \citet{Herreman2015} who refer by it to all pinch-type instabilities
studied by R. J. Tayler including also the classical case with deformable
boundaries \citep{Tayler1957,Tayler1960}. It is important to note
that surface deformation provides an additional pinch instability
mechanism. Although surface deformation always involves fluid flow,
it is not required for the internal pinch-type instability, which
can be driven by the fluid flow alone. In ideally conducting fluid,
where the magnetic field is frozen in, flow causes the same-order
disturbance of the magnetic field as that associated with the surface
deformation, i.e., proportional to the displacement \citep{Tayler1957}.
In a poorly conducting liquid, the perturbation of the magnetic field
caused by the fluid flow is proportional not to the displacement but
to the product of velocity and conductivity. This perturbation is
much weaker than that associated with surface deformation \citep{Tayler1960}.

Another peculiarity of internal current-driven instability is its
reliance on the curvature of the magnetic field. For example, in the
liquid metal annulus carrying an axial current and bounded by solid
walls, instability vanishes when the radius of the gap tends to infinity
and, thus, the circular magnetic flux lines straighten out \citep{Priede2015}.
This is consistent with the absence of purely magnetic instability
in the planar perfectly conducting liquid layer permeated by a straight
co-planar magnetic field whatever its distribution over the thickness
of the layer \citep{Ogilvie1996}. In contrast to cylindrical geometries,
where the curvature of the magnetic field is inherent and pinch-type
instabilities usually occur, in planar geometries, such instabilities
are likely to be induced by the edges and corners, around which the
magnetic field bends. Moreover, rectangular configurations may be
relevant to the recently developed liquid metal batteries \citep{Wang2014},
for which the electrode mixing \citep{Kelley2014} due to potential
current-driven instabilities is one of the major concerns \citep{Stefani2011,Weber2013,Weber2015,Herreman2015}.

In this paper we consider an elementary model of pinch-type instability
consisting of a semi-infinite planar layer of inviscid conducting
liquid that is bounded by solid walls and carries a uniform electric
current. This is the simplest possible model of the internal pinch
instability in planar geometry. It allows us to elucidate the basic
characteristics of the instability, which may otherwise be obscured
by technical details of more realistic numerical models and laboratory
experiments. In particular, we show that the internal pinch instability
is an edge effect caused by the curvature of the magnetic field, and
its growth rate is determined by the linear current density and independent
of the system size. This suggests that the instability can be prevented
either by reducing the field curvature at the edges or by limiting
the layer thickness but not its lateral size when the areal current
density is fixed. We also point out that the instability is significantly
different in well conducting and highly resistive fluids like liquid
metals. In the latter, the instability development time depends on
the electrical conductivity rather than being determined by the Alfvén
wave speed as for the Tayler instability. This somewhat limits the
astrophysical significance of liquid-metal laboratory experiments.
On the other hand, it implies that this type of instability, if any,
in the liquid metal batteries significantly differs from the Tayler
instability in astrophysics.

The paper is organized as follows. Mathematical model is introduced
in $\S$\ref{sec:prob} and the linear stability problem is formulated
in $\S$\ref{sec:lin-stab}. Numerical method and the main results
are presented in $\S$\ref{sec:Numeth} and $\S$\ref{sec:Res}. The
paper is concluded with a summary and brief discussion of the results
in $\S$\ref{sec:Sum}.

\section{\label{sec:prob}Formulation of the problem}

\begin{figure}
\begin{centering}
\includegraphics[bb=0bp 0bp 213bp 179bp,width=0.5\textwidth]{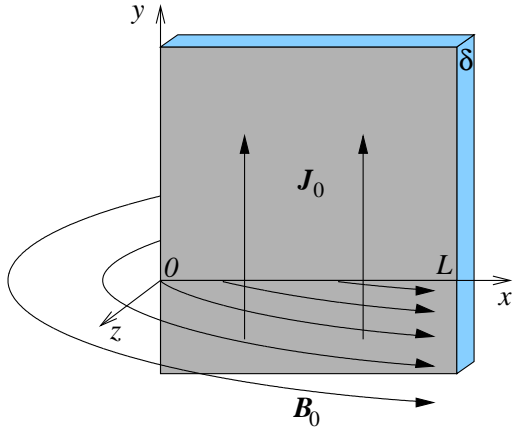}
\par\end{centering}
\caption{\label{fig:sketch}Sketch of the problem.}
\end{figure}

Consider a channel of width $L$ and depth $\delta$ between two solid
vertical parallel walls filled with an inviscid liquid of density
$\rho$ and electrical conductivity $\sigma.$ The liquid carries
electric current of a uniform density $\vec{j}_{0}$ directed upwards
along the rigid edge of the channel, which coincides with the $y$-axis
of the Cartesian system of co-ordinates shown in figure \ref{fig:sketch}.
The layer is assumed to be thin with $\delta\ll L$ and, thus, treated
as a planar undeformable sheet with the areal density $\bar{\rho}=\rho\delta$
and the respective electrical conductivity $\bar{\sigma}=\sigma\delta.$
The electric current with the linear density $\vec{J}_{0}=\vec{e}_{y}j_{0}\delta$
that flows along the sheet generates a magnetic field whose normal
$(z)$ component through the sheet can be found using the Biot-Savart
law as
\begin{equation}
B_{0}(x)=\frac{\mu_{0}J_{0}}{2\pi}\int_{0}^{L}\frac{d\xi}{\xi-x}=\frac{\mu_{0}J_{0}}{2\pi}\ln\left(\frac{L}{x}-1\right)\approx\frac{\mu_{0}J_{0}}{2\pi}\ln\frac{L}{x},\label{eq:B0}
\end{equation}
where the last approximate expression holds in the vicinity of edge
for $x\ll L,$ which will be the main focus of this study. On the
other hand, the sheet approximation implies $x\gg\delta.$ Note that
expression (\ref{eq:B0}) represents a superposition of the fields
of straight wires which may be thought to form the sheet. As only
the gradient of the field $B_{0}'(x)\approx-\frac{\mu_{0}J_{0}}{2\pi x}$
is required in the following, the width $L$ drops out of the model.
It means that the layer is treated as semi-infinite. The interaction
of the electric current with its own magnetic field gives rise to
the pinch force with the areal density $\vec{F}_{0}=\vec{J}_{0}\times\vec{B}_{0}=\vec{e}_{x}J_{0}B_{0}(x),$
which is balanced by the $x$-component of the effective pressure
gradient
\begin{equation}
P_{0}'(x)=-J_{0}B_{0}(x).\label{eq:p0}
\end{equation}
Note that for sheet, the effective pressure $P$ is defined as force
per unit length.

Disturbance of this quiescent equilibrium state by an arbitrary slow
liquid flow, which is confined to the plane of the sheet by the solid
bounding walls, is governed by the linearised 2D Euler's equation
\begin{equation}
\bar{\rho}\partial_{t}\vec{v}_{1}=-\vec{\nabla}P_{1}+\vec{F}_{1}\label{eq:Euler}
\end{equation}
with the velocity distribution $\vec{v}_{1}$ subject to the incompressibility
constraint $\vec{\nabla}\cdot\vec{v}_{1}=0.$ The linearised perturbation
of the electromagnetic force $\vec{F}_{1}=(\vec{J}_{1}B_{0}+\vec{J}_{0}B_{1})\times\vec{e}_{z}$
produced by the flow results from the interaction of the current perturbation
$\vec{J}_{1}$ with the base magnetic field $B_{0}$ as well as from
the interaction of the base current $\vec{J}_{0}$ with the perturbation
of the normal magnetic field $B_{1}.$ Note that the quadratic terms
in the perturbation amplitudes are neglected as usual in the linear
stability analysis because all perturbations are assumed to be infinitesimal.
The perturbation of the pressure gradient $\vec{\nabla}P_{1}$ is
eliminated from equation (\ref{eq:Euler}) by applying the\emph{ curl}
operator. Then the flow is governed by the $z$-component of the resulting
equation 
\begin{equation}
\bar{\rho}\partial_{t}\omega_{1}=-(\vec{J}_{1}\cdot\vec{\nabla}B_{0}+\vec{J}_{0}\cdot\vec{\nabla}B_{1}),\label{eq:vrtc}
\end{equation}
where $\omega_{1}=\vec{e}_{z}\cdot\vec{\nabla}\times\vec{v}_{1}$
is the normal component of vorticity. The associated current perturbation
is governed by Ohm's law for a moving medium 
\begin{equation}
\vec{J}_{1}=\bar{\sigma}(\vec{E}_{1}+\vec{v}_{1}\times\vec{e}_{z}B_{0}).\label{eq:Ohm}
\end{equation}
 A time-dependent perturbation of the magnetic field induces a rotational
electric field according to the first Maxwell equation $\vec{\nabla}\times\vec{E}_{1}=-\partial_{t}\vec{B}_{1}.$
Note that in highly resistive media this induction effect is usually
negligible, which corresponds to the so-called quasi-stationary or
inductionless approximation commonly used in the liquid-metal MHD
\citep{Roberts1967}. In order to keep our model general and applicable
also to well conducting fluids, we forgo this approximation here.

The charge conservation $\vec{\nabla}\cdot\vec{J}_{1}=0$ is satisfied
by introducing the electric stream function $h_{1},$ which, as shown
below, is directly related to the scalar magnetic potential at the
current sheet. It allows us to represent the electric current in the
plane of sheet as 
\begin{equation}
\vec{J}_{1}=-\vec{e}_{z}\times\vec{\nabla}h_{1}.\label{eq:J1}
\end{equation}
Note that a truly 2D ($z$-independent) current distribution would
generate a 2D magnetic field with $z$-component only which would
then coincide with the current stream function introduced above. The
current sheet considered here, however, produces a 3D magnetic field
whose normal component at the sheet is related with the electric stream
function by the induction-type equation which is obtained as follows.
Substituting this into equation (\ref{eq:Ohm}) and taking the $z$-component
of the \emph{curl} of the resulting equation, we obtain 
\begin{equation}
\partial_{t}B_{1}+\vec{v}_{1}\cdot\vec{\nabla}B_{0}=\bar{\sigma}^{-1}\vec{\nabla}^{2}h_{1}.\label{eq:h}
\end{equation}
The magnetic field perturbation in the surrounding space is sought
as 
\begin{equation}
\vec{B}_{1}=-\mu_{0}\vec{\nabla}\Psi_{1}\label{eq:B1}
\end{equation}
 using the scalar magnetic potential $\Psi_{1},$ which is governed
by the Laplace equation 
\begin{equation}
\vec{\nabla}^{2}\Psi_{1}=0.\label{eq:psi}
\end{equation}
The surface current density is related to the jump of the tangential
magnetic field over the current sheet by Ampère's circuital law: $\vec{J}_{1}=\frac{1}{\mu_{0}}\vec{e}_{z}\times\left[\vec{B}_{1}\right]_{S},$
where $\mu_{0}=\unit[4\pi\times10^{-7}]{H/m}$ is the vacuum permeability.
Combining this current sheet condition with equations (\ref{eq:J1})
and (\ref{eq:B1}), we obtain the boundary condition 
\begin{equation}
\left.\Psi_{1}\right|_{z\rightarrow+0}=-\left.\Psi_{1}\right|_{z\rightarrow-0}=\frac{1}{2}h_{1},\label{bc:psi}
\end{equation}
which relates the scalar magnetic potential with the electric stream
function. It is important to note that the problem is defined in terms
of the linear current density $J_{0}$ which is the relevant electromagnetic
parameter in the current sheet approximation. 

\section{\label{sec:lin-stab}Linear stability analysis}

Owing to the $y$-invariance of the base state, perturbations can
be sought in the normal mode form 
\[
\{\vec{v}_{1},p_{1},h_{1}\}(\vec{x},t)=\{\vec{\hat{v}},\hat{p},\hat{h}\}(x)e^{\gamma t+iky},
\]
where $\gamma$ is a generally complex growth rate and $k$ is a real
wavenumber. Then equations (\ref{eq:vrtc}) and (\ref{eq:h}) for
the flow and current perturbations take the form
\begin{eqnarray}
\gamma\vec{D}_{k}^{2}\hat{v} & = & -\bar{\rho}^{-1}J_{0}k^{2}\Bigl(\hat{B}+\frac{\mu_{0}}{2\pi}\frac{\hat{h}}{x}\Bigr),\label{eq:vhh}\\
\gamma\hat{B} & = & \bar{\sigma}^{-1}\vec{D}_{k}^{2}\hat{h}+J_{0}\frac{\mu_{0}}{2\pi}\frac{\hat{v}}{x},\label{eq:hhv}
\end{eqnarray}
where $\hat{v}$ is the $x$-component of the velocity perturbation,
which is further referred to as the transverse velocity, $\hat{B}$
is the normal $(z)$ component of the induced magnetic field at the
sheet, and $\vec{D}_{k}\equiv\vec{e}_{x}\frac{\mathrm{d}\,}{\mathrm{d}x}+ik\vec{e}_{y}$
is a spectral counterpart of the nabla operator acting on the mode
with the wavenumber $k$. The $y$-component of velocity perturbation
$\hat{u}$ has been eliminated from (\ref{eq:vhh}) by using the incompressibility
constraint $\vec{D}_{k}\cdot\vec{\hat{v}}=\hat{v}'+ik\hat{u}=0,$
while the $z$-component of velocity is absent owing to the undeformability
of the sheet.

The boundary conditions which follow from the vanishing of the $x$-component
of velocity and electric current at the fixed insulating edge $(x=0)$
and far from the edge $(x\rightarrow\infty)$ are 
\begin{equation}
\hat{v}=\hat{h}=0.\label{eq:bc}
\end{equation}
These boundary conditions imply $(\hat{v},\hat{h})\sim x$ at the
edge. It means that the terms $\sim1/x,$ which result from the logarithmic
singularity of the base magnetic field (\ref{eq:B0}) in equations
(\ref{eq:vhh}) and (\ref{eq:hhv}), are expected to be regular as
$x\rightarrow0:$ $\hat{h}/x\rightarrow\hat{h}'(0)$ and $\hat{v}/x\rightarrow\hat{v}'(0).$
Then the finite longitudinal current perturbation at the edge $\hat{h}'(0)=\hat{J}_{y}(0)$
would produce a perturbation of the magnetic field with a logarithmic
singularity $\hat{B}\sim\ln x$ similar to that of the base magnetic
field (\ref{eq:B0}). According to equations (\ref{eq:vhh}) and (\ref{eq:hhv}),
which reduce to $(\hat{v},\hat{h})''\sim\ln x$ for $x\rightarrow0$,
logarithmic singularity produces a higher order small perturbation
$(\hat{v},\hat{h})\sim x^{2}\ln x.$ This confirms the regularity
of $\hat{h}'$ and $\hat{v}'$at the edge. 

To find $\hat{B}$ required in equation (\ref{eq:vhh}) we need to
solve equation (\ref{eq:psi}) in the space surrounding the sheet.
This can conveniently be done in the cylindrical coordinates with
the axis aligned along the edge and the poloidal angle $\theta$ measured
from the plane of the sheet so that $x=r\cos\theta,$ $z=r\sin\theta$
and $\Psi_{1}(\vec{x},t)=\hat{\Psi}(r,\theta)e^{\gamma t+iky}.$ Then
equation (\ref{eq:psi}) takes the form 
\begin{equation}
r\partial_{r}(r\partial_{r}\hat{\Psi})+\partial_{\theta}^{2}\hat{\Psi}-(rk)^{2}\hat{\Psi}=0.\label{eq:ph}
\end{equation}
Boundary condition (\ref{bc:psi}), which now reads as $\left.\hat{\Psi}\right|_{\theta=0}=-\left.\hat{\Psi}\right|_{\theta=2\pi}=\frac{1}{2}\hat{h}(x),$
and the reflection symmetry of the problem $\hat{\Psi}(r,\theta)=-\hat{\Psi}(r,2\pi-\theta),$
suggest a solution in the form 
\begin{equation}
\hat{\Psi}(r,\theta)=\frac{1}{2}\hat{h}(r)\cos\frac{\theta}{2}-\frac{1}{2\pi}\sum_{n=1}^{\infty}c_{n}\hat{\Psi}_{n}(r)\sin(n\theta),\label{eq:psih}
\end{equation}
where $c_{n}=\int_{0}^{2\pi}\sin(n\theta)\cos\frac{\theta}{2}\,\mathrm{d}\theta=\frac{8n}{4n^{2}-1}.$
Substituting this expression into equation (\ref{eq:ph}), we obtain
a sequence of ODE's 
\begin{equation}
r(r\hat{\Psi}_{n}')'-(n^{2}+(rk)^{2})\hat{\Psi}_{n}=r(r\hat{h}')'-(2^{-2}+(rk)^{2})\hat{h},\label{eq:phh}
\end{equation}
which define the harmonics of the scalar magnetic potential $\hat{\Psi}_{n}$
in terms of the electric stream function $\hat{h}.$ For (\ref{eq:psih})
to be single valued at $r=0,$ $\hat{\Psi}_{n}(0)=0$ is required,
which serves as a boundary condition for equation (\ref{eq:phh}).
Then the normal magnetic field perturbation in equation (\ref{eq:vhh})
is given by 
\begin{equation}
\hat{B}(x)=-\frac{\mu_{0}}{x}\left.\partial_{\theta}\hat{\Psi}\right|_{\theta=0}=\frac{\mu_{0}}{2\pi x}\sum_{n=1}^{\infty}nc_{n}\hat{\Psi}_{n}(x).\label{eq:H}
\end{equation}

The problem is converted into the dimensionless form by using $v_{m}=(\mu_{0}\bar{\sigma})^{-1}$
and 
\begin{equation}
\tau_{m}=\frac{\bar{\rho}}{\bar{\sigma}\mu_{0}^{2}J_{0}^{2}}\label{eq:tau-m}
\end{equation}
as the characteristic velocity and time scales. The former corresponds
to the magnetic diffusion speed, whereas the latter, when represented
in terms of the characteristic magnetic flux density $B_{0}=\mu_{0}J_{0},$
coincides with the so-called magnetic response time \citep{Roberts1967},
also known as the magnetic damping time \citep{Davidson2001}, which
typically appears in the inductionless \global\long\def\Pm{\mathrm{Pm}}
$(\Pm=0)$ limit. The length scale can conveniently be chosen as 
\begin{equation}
k^{-1}=\frac{\lambda}{2\pi},\label{eq:lmb}
\end{equation}
where $k$ and $\lambda$ are the wavenumber and wave length of perturbation,
respectively. The electric stream function and the scalar magnetic
potential are both scaled with $J_{0}/k.$ Then equations (\ref{eq:hhv})
and (\ref{eq:vhh}) take the following dimensionless form\global\long\def\Sl{\mathrm{S}_{\lambda}}
 
\begin{eqnarray}
\gamma\vec{D}_{k}^{2}\hat{v} & = & -\Bigl(\hat{B}+\frac{1}{2\pi}\frac{\hat{h}}{x}\Bigr),\label{eq:vhn}\\
\gamma\Sl^{2}\hat{B} & = & \vec{D}_{k}^{2}\hat{h}+\frac{1}{2\pi}\frac{\hat{v}}{x},\label{eq:hhn}
\end{eqnarray}
where the dimensionless wavenumber $k=1$ and $\Sl=\bar{\sigma}\mu_{0}J_{0}\sqrt{\frac{\mu_{0}\lambda}{2\pi\bar{\rho}}}=\frac{v_{A}}{v_{m}}$
is the Lundquist number based on the wave length $\lambda.$ This
is the only parameter of the problem, which defines the Alfvén speed
$v_{A}=J_{0}\sqrt{\frac{\mu_{0}\lambda}{2\pi\bar{\rho}}}$ relative
to that of the magnetic diffusion $v_{m}$ introduced above.  A small
$\Sl$ corresponds to the quasi-stationary limit in which the induction
effect represented by the l.h.s. term of equation (\ref{eq:hhn})
becomes negligible. In the limit of vanishing Lundquist number $(\Sl=0),$
the growth rate $\gamma$ becomes independent of $\Sl.$ In the opposite
limit $(\Sl\gg1),$ the magnetic diffusion represented by the first
term on the RHS of equation (\ref{eq:hhn}) is expected to become
negligible. In this case, which corresponds to an ideally conducting
liquid, $\hat{v}$ can be substituted from equation (\ref{eq:hhn})
into equation (\ref{eq:vhn}). Then $\gamma^{2}\Sl^{2}$ emerges as
the only parameter (eigenvalue) of the reduced problem. Consequently,
in this limit, we expect $\gamma\sim\Sl^{-1}.$ The respective physical
instability development time is 
\begin{equation}
\frac{\tau_{m}}{\gamma}\sim J_{0}^{-1}\sqrt{\frac{\bar{\rho}\lambda}{2\pi\mu_{0}}}=\frac{1}{2\pi}\frac{\lambda}{v_{A}},\label{eq:t-alfv}
\end{equation}
which is the characteristic Alfvén time for the length scale (\ref{eq:lmb}). 

\section{\label{sec:Numeth}Numerical method}

The eigenvalue problem posed by equations (\ref{eq:vhn}), (\ref{eq:hhn})
and (\ref{eq:phh}) was solved numerically using a Chebyshev collocation
method with Chebyshev-Lobatto nodes $\eta_{m}=\cos\left(m\pi/(M+1)\right),\quad m=0,\cdots,M+1,$
and the coordinate transform $\eta=\frac{\alpha x-1}{\alpha x+1}$
that maps the semi-infinite domain $x=r\in[0,\infty)$ onto $\eta=[-1,1]$
using parameter $\alpha$ and transforms the differentiation operator
$\frac{\mathrm{d}\,}{\mathrm{d}x}$ into $\frac{1}{2}\alpha(1-\eta)^{2}\frac{\mathrm{d}\,}{\mathrm{d}\eta}.$
Equations were approximated at the internal collocation points $x_{m}$
corresponding to $m=1,\cdots M$ and the boundary conditions were
applied at $x_{M+1}=0$ and $x_{0}=\infty$ \citep{Boyd2013}. This
eliminates the potential edge singularity of the normal magnetic field
from the discretizied problem. The problem was reduced to a standard
matrix eigenvalue problem as follows. First, the Fourier series (\ref{eq:psih})
was truncated at the length $N.$ Second, the normal magnetic field
$\hat{B}$ defined by equation (\ref{eq:H}) was expressed in terms
of the electric stream function $\hat{h}$ by inverting $N$ matrices
of size $M\times M$ representing the l.h.s. operators in (\ref{eq:phh})
for $n=1,\cdots,N$ Fourier modes, and then substituted into equations
(\ref{eq:vhn}) and (\ref{eq:hhn}). Eventually, inverting the matrix
representations of the l.h.s operators of the resulting equations,
we obtained a standard matrix eigenvalue problem of size $2M\times2M$
for the growth rate $\gamma$ and the unknown vector $(\hat{v},\hat{h})$
at the internal collocation points. In the quasi-stationary limit,
which corresponds to $\Sl=0,$ the problem can be simplified further
by expressing $\hat{h}$ from equation (\ref{eq:hhn}) in terms of
$\hat{v}$ and then substituting it into equation (\ref{eq:vhn}).
This results in a standard matrix eigenvalue problem of size $M\times M$
for the growth rate $\gamma$ and the vector of $\hat{v}.$ In the
ideally conducting limit, which corresponds to $S_{\lambda}\gg1,$
the problem is reduced to a standard matrix eigenvalue problem of
the same size $M\times M$ for the eigenvalue $\gamma^{2}\Sl^{2}$
and the vector $\hat{h}$ as outlined in the previous section. Matrices
were inverted and eigenvalue problem solved using the \emph{LU} and
\emph{QR} factorization algorithms from the standard linear algebra
software library LAPACK. 

\section{\label{sec:Res}Results}

\begin{figure}
\begin{centering}
\includegraphics[width=0.5\textwidth]{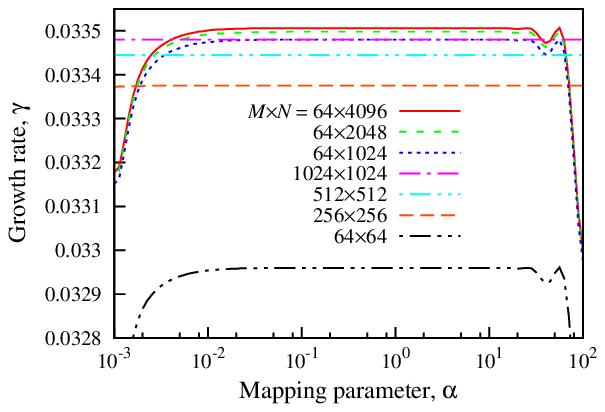}\put(-30,30){(a)}\includegraphics[width=0.5\textwidth]{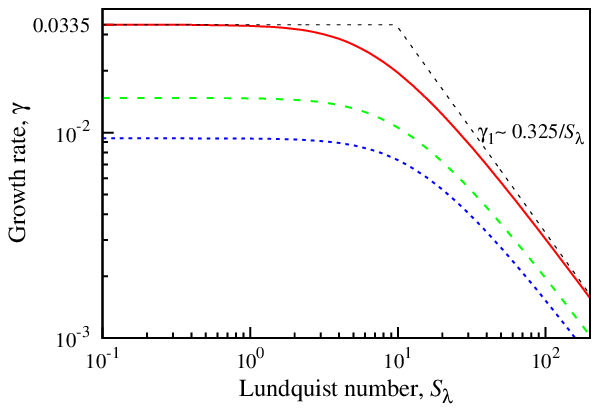}\put(-160,30){(b)}
\par\end{centering}
\caption{\label{fig:grate}Growth rate of the leading quasi-stationary $(\protect\Sl=0)$
mode computed with various number of collocation points $M$ and Fourier
modes $N$ versus the mapping parameter $\alpha$ (a); growth rates
of the three leading modes versus the Lundquist number computed with
$M=64,N=4096,\alpha=0.3$ (b). }
\end{figure}

We start with the quasi-stationary limit, which corresponds to $\Sl=0,$
and first verify the accuracy of our numerical method. The highest
growth rate $\gamma_{1}$ computed in this limit with various numbers
of collocation points $M$ and Fourier modes $N$ is plotted in figure
\ref{fig:grate}(a) against the mapping parameter $\alpha.$ First,
note that $\gamma_{1}$ is real and positive, which means a monotonic
instability. The same holds also for the two subsequent eigenvalues
shown in table \ref{tab:gamma}, which are several times smaller than
$\gamma_{1}.$ Second, the growth rate is seen to be nearly constant
over a wide range of the mapping parameter $\alpha.$ For $M=64$
collocation points, a noticeable variation, which is a purely numerical
effect, appears only for $\alpha\lesssim10^{-2}$ and $\alpha\gtrsim10.$
The growth rate computed with $M=64$ is nearly constant almost over
three decades of $\alpha.$ As seen in figure \ref{fig:grate}, the
range of constant $\gamma_{1}$ extends over more than five decades
of $\alpha$ when $M\ge256.$ In the following, we fix $\alpha=0.3,$
which is close to the centre of this range. Then $M=64$ produces
$\gamma_{1}$ coinciding up four decimal places with the value computed
with $M=1024,$ which is a typically fast convergence of the Chebyshev
collocation approximation. The convergence, however, is much slower
with respect to the number of Fourier modes $N$. As seen in table
\ref{tab:gamma}, $N\gtrsim1024$ is required to compute the leading
eigenvalue $\gamma_{1}\approx0.0335$ with 4 d.p. Convergence is presumably
slowed down by the edge singularity, which is discussed below, as
well as by the discontinuity of the scalar magnetic potential over
the current sheet resulting from the boundary condition (\ref{bc:psi}).
This discontinuity limits the Fourier series convergence rate to algebraic
while the convergence rate of the Chebyshev collocation approximation
is typically exponential \citep{Gottlieb1977}. Further, we use $N=4096$
and $M=64.$ 

\begin{table}
\begin{centering}
\begin{tabular}{ccccc}
$M$ & $N$ & $\gamma_{1}$ & $\gamma_{2}$ & $\gamma_{3}$\tabularnewline
$64$ & $64$ & $0.032960$ & $0.014264$ & $0.0089024$\tabularnewline
$256$ & $256$ & $0.033375$ & $0.014653$ & $0.0092855$\tabularnewline
$1024$ & $1024$ & $0.033480$ & $0.014752$ & $0.0093836$\tabularnewline
$64$ & $1024$ & $0.033480$ & $0.014752$ & $0.0093836$\tabularnewline
$64$ & $2048$ & $0.033498$ & $0.014769$ & $0.0094000$\tabularnewline
$64$ & $4096$ & $0.033506$ & $0.014777$ & $0.0094082$\tabularnewline
\end{tabular}
\par\end{centering}
\caption{\label{tab:gamma}Growth rates of the three leading quasi-stationary
$(S_{\lambda}=0)$ modes computed with various number collocation
points $M$, Fourier modes $N$ and $\alpha=0.3.$}
\end{table}

\begin{figure}
\begin{centering}
\includegraphics[bb=50bp 50bp 230bp 176bp,width=0.5\columnwidth]{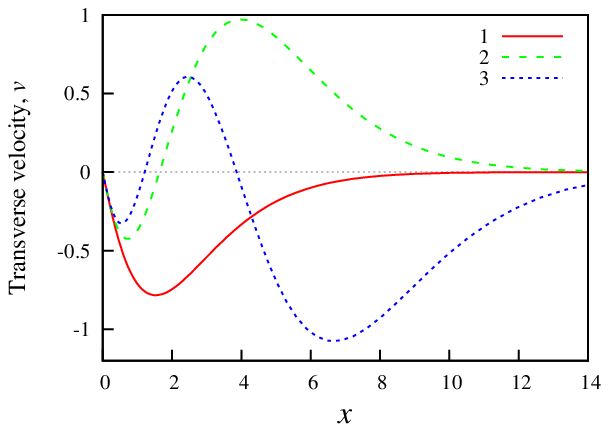}\put(-25,25){(a)}\includegraphics[bb=50bp 50bp 230bp 176bp,width=0.5\columnwidth]{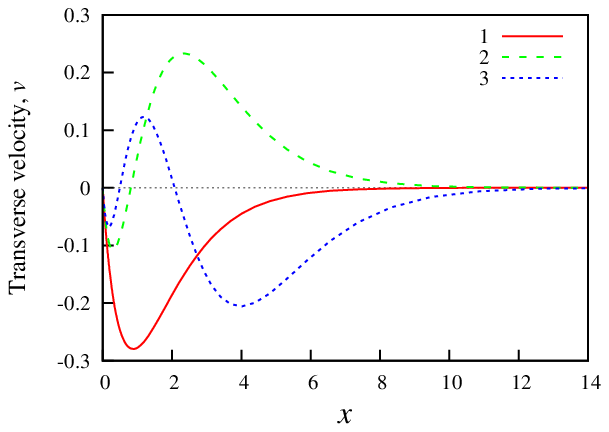}\put(-25,25){(b)}
\par\end{centering}
\begin{centering}
\includegraphics[width=0.5\columnwidth]{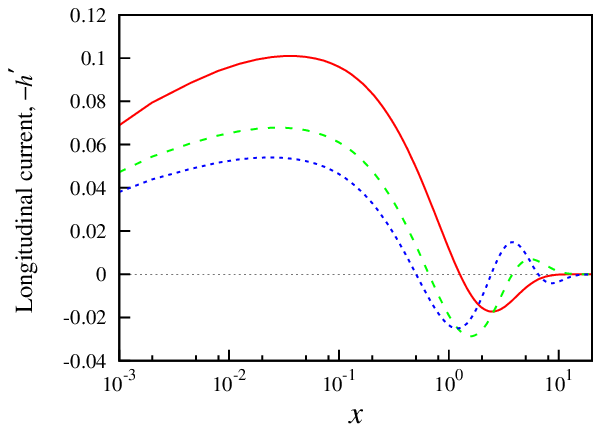}\put(-25,25){(c)}\includegraphics[width=0.5\columnwidth]{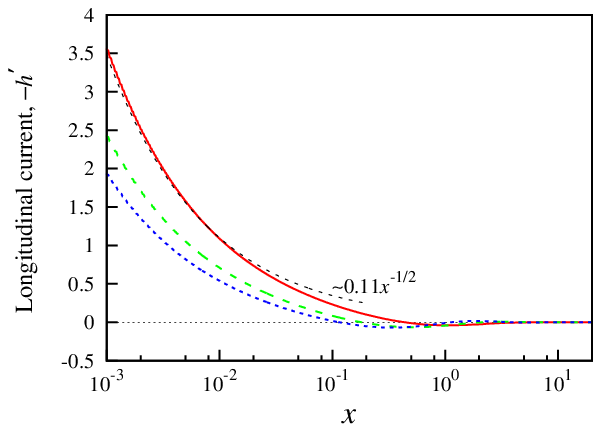}\put(-25,25){(d)}
\par\end{centering}
\begin{centering}
\includegraphics[bb=50bp 50bp 230bp 176bp,clip,width=0.5\columnwidth]{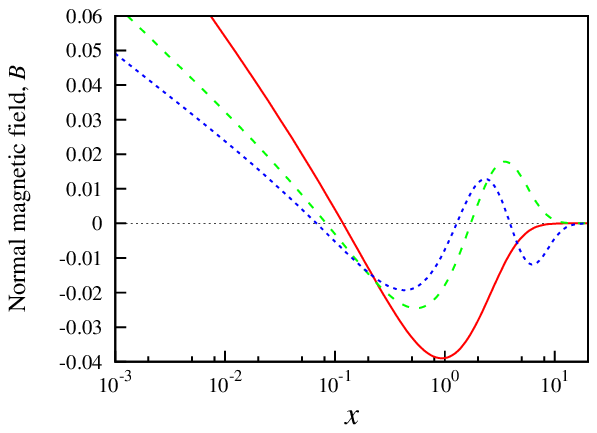}\put(-25,25){(e)}\includegraphics[bb=50bp 50bp 230bp 176bp,clip,width=0.5\columnwidth]{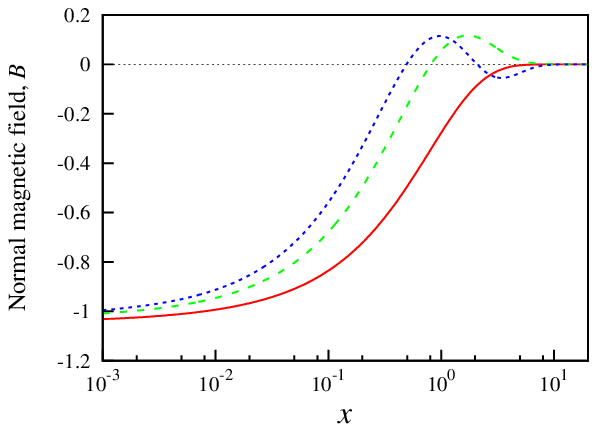}\put(-25,25){(f)}
\par\end{centering}
\caption{\label{fig:egv-x}Amplitude distributions of the transverse velocity
$\hat{v}$ (a,b), the longitudinal current density $-\hat{h}'$ (c,d)
and the normal magnetic field $\hat{B}$ (e,f) for the three fastest
growing modes in the quasi-stationary (a,c,e) and ideally conducting
(b,d,f) limits. The amplitudes are normalized by the condition $\hat{v}'(0)=-1.$
The normal magnetic field in the ideally conducting case is rescaled
with $(2\pi\gamma\protect\Sl^{2})^{-1}$ so that $\hat{B}(0)=\hat{v}'(0).$ }
\end{figure}

Growth rates of the three leading modes computed using the full dynamical
model (\ref{eq:vhn}, \ref{eq:hhn}) are plotted in figure \ref{fig:grate}(b)
versus the Lundquist number $\Sl.$ For $\Sl\ll1,$ the quasi-stationary
limit considered above is obviously recovered. As $\Sl$ increases,
the development of instability is slowed down by the increasing magnetic
diffusion time. For $\Sl\gtrsim10,$ the growth rate is seen to reduce
as $\gamma\sim\Sl^{-1},$ which confirms the asymptotics inferred
in $\S$\ref{sec:lin-stab}. The asymptotic growth rate of the dominant
$\gamma\sim0.325/\Sl$ is seen to agree well with the prediction of
the reduced dynamical model for ideally conducting liquid described
in the previous section.

\begin{figure}
\begin{centering}
\includegraphics[bb=0bp 0bp 242bp 238bp,width=0.33\textwidth]{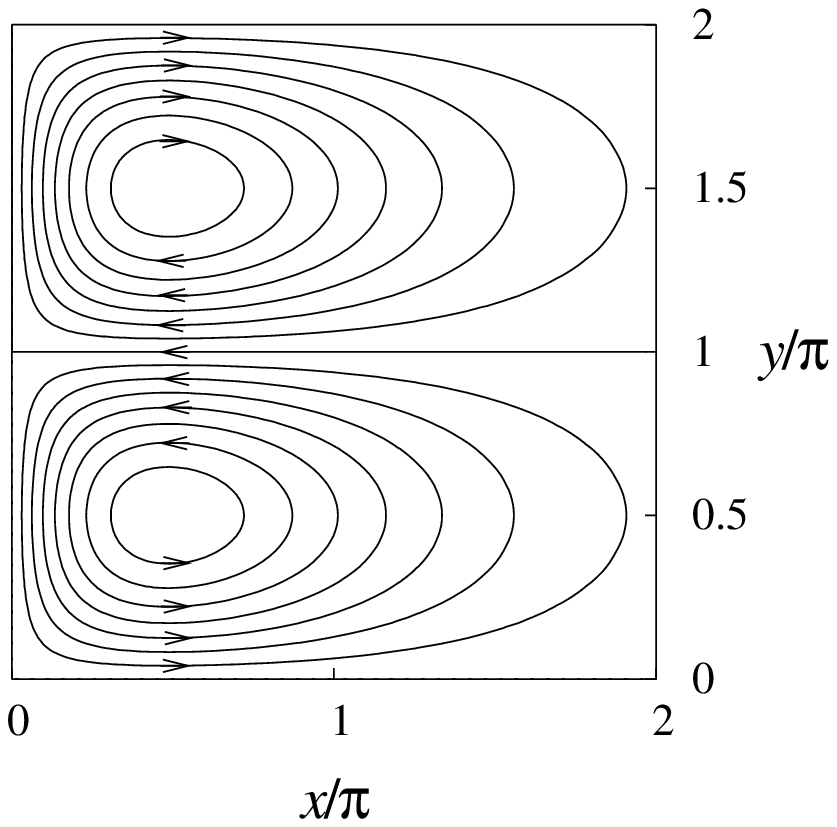}\put(-42,25){(a)}\includegraphics[bb=0bp 0bp 242bp 238bp,width=0.33\textwidth]{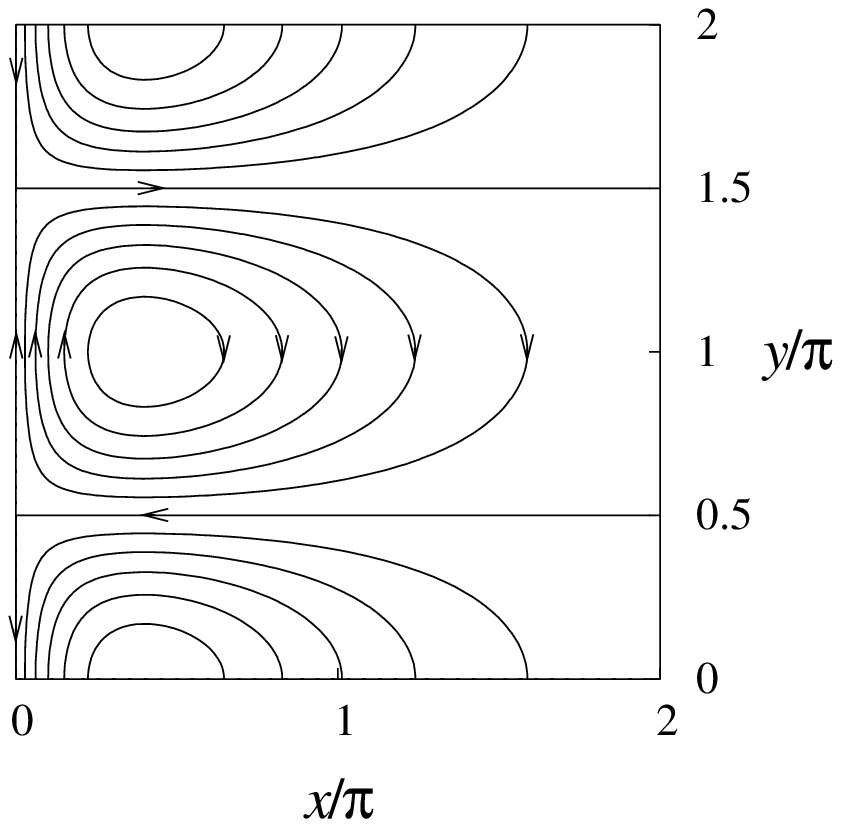}\put(-42,25){(b)}\includegraphics[width=0.33\textwidth]{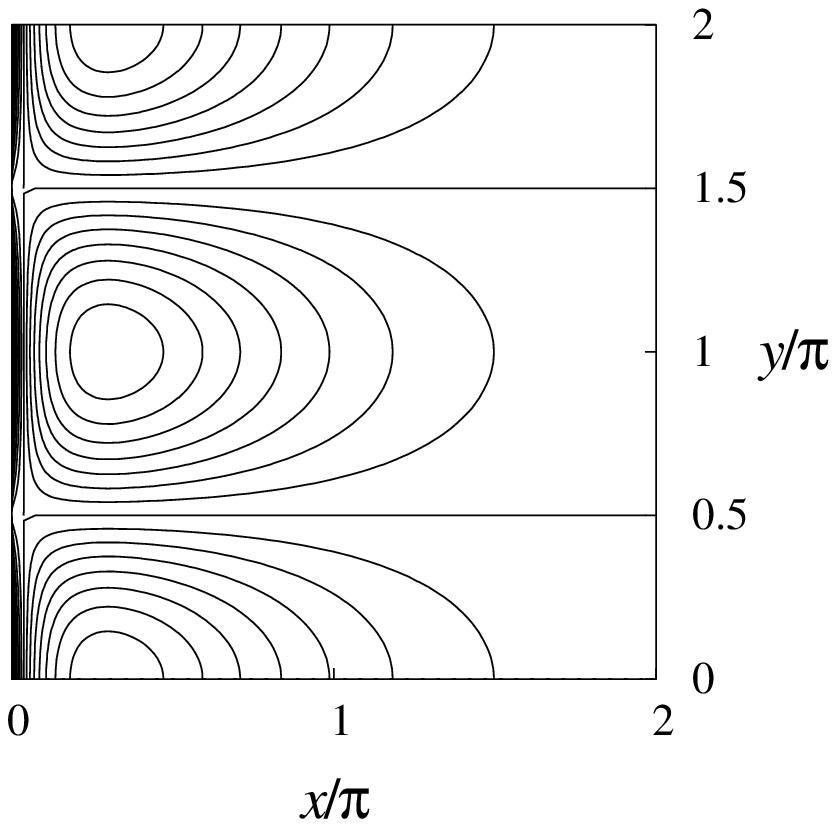}\put(-42,25){(c)}
\par\end{centering}
\begin{centering}
\includegraphics[bb=0bp 0bp 242bp 238bp,width=0.33\textwidth]{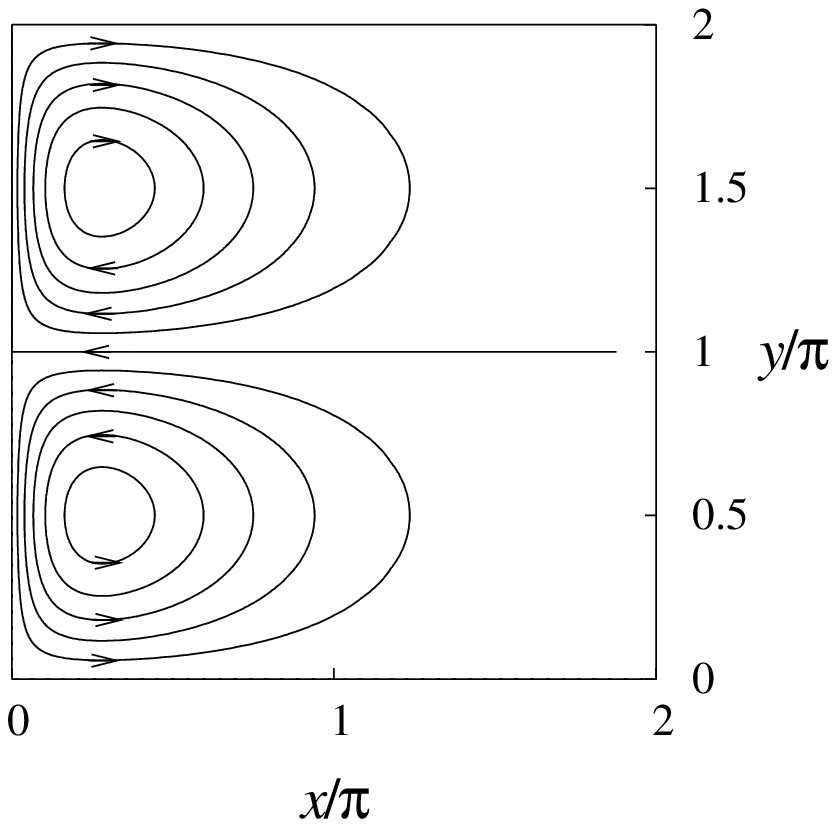}\put(-42,25){(d)}\includegraphics[bb=0bp 0bp 242bp 238bp,width=0.33\textwidth]{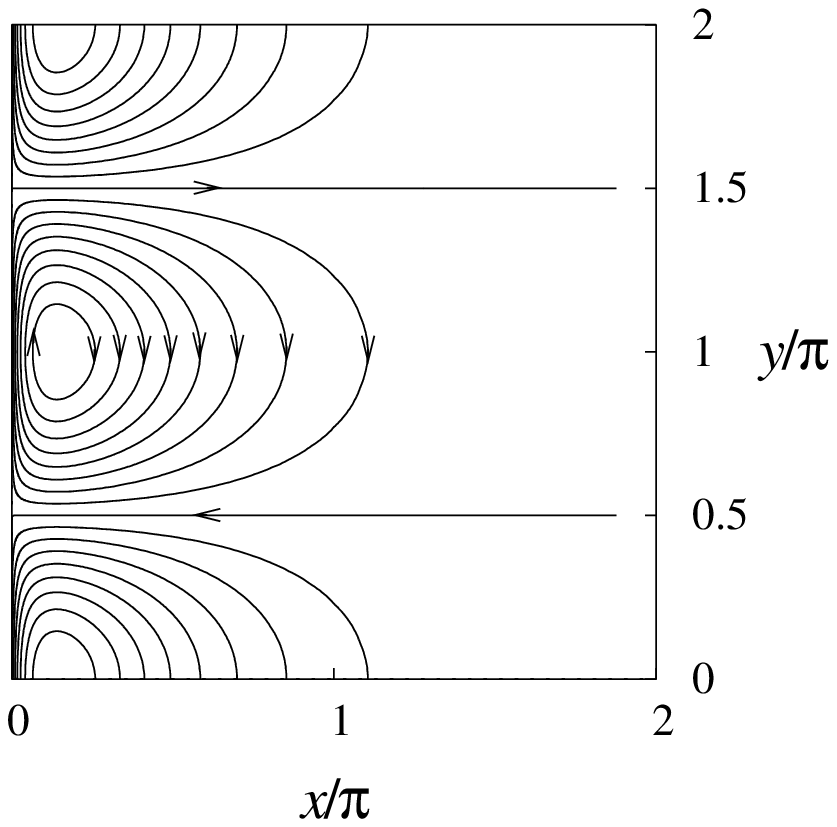}\put(-42,25){(e)}\includegraphics[bb=0bp 0bp 242bp 238bp,width=0.33\textwidth]{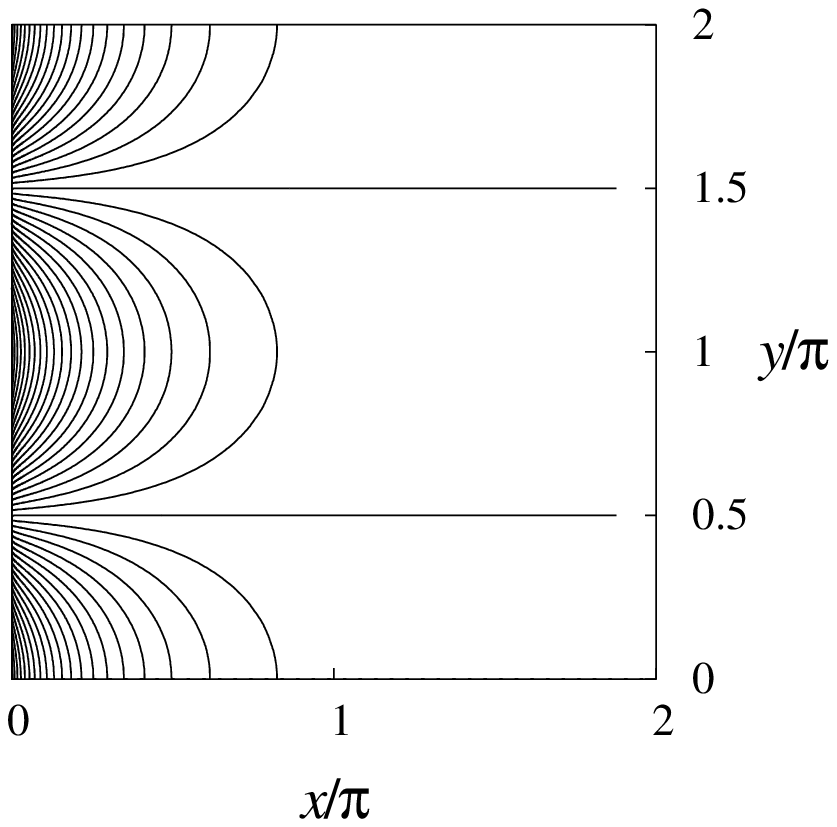}\put(-42,25){(f)}
\par\end{centering}
\caption{\label{fig:egv-xy}Streamlines (a,d), electric current lines (b,e)
and the normal magnetic field isolines (c,f) of the fastest growing
mode in the quasi-stationary (a,b,c) and ideally conducting (d,e,f)
limits.}
\end{figure}

\begin{figure}
\begin{centering}
\includegraphics[bb=120bp 90bp 380bp 280bp,clip,width=0.5\columnwidth]{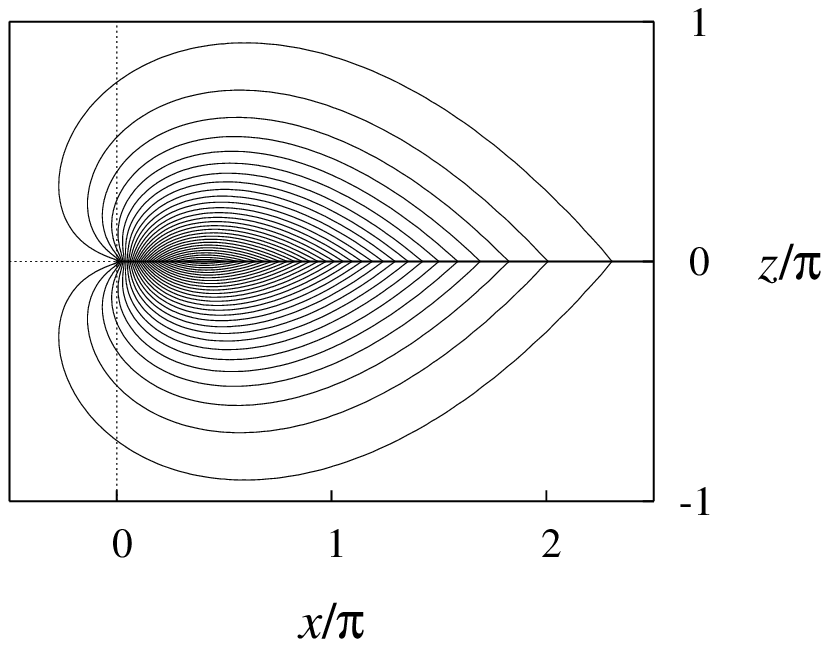}\put(-60,35){(a)}\includegraphics[bb=120bp 90bp 380bp 280bp,clip,width=0.5\columnwidth]{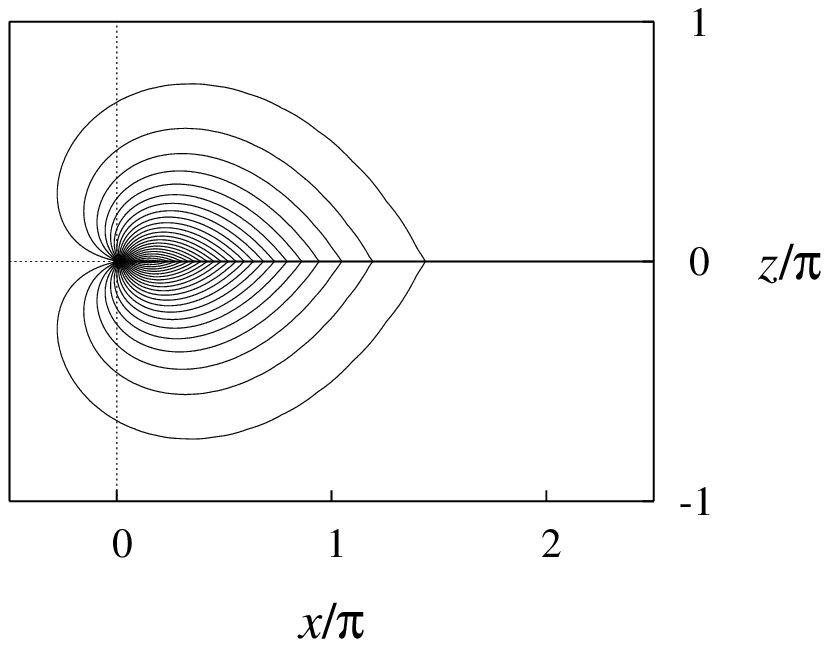}\put(-60,35){(b)}
\par\end{centering}
\caption{\label{fig:psi-xz}Amplitude distributions of the scalar magnetic
potential perturbation $\hat{\Psi}$ in the $(x,z)$-plane at the
edge of sheet for the fastest growing mode in the quasi-stationary
(a) and ideally conducting (b) limits.}
\end{figure}

Amplitude distributions of the three fastest growing modes in both
poorly and ideally conducting limits are plotted versus the distance
from the edge in figure \ref{fig:egv-x}. The fastest growing mode,
whose streamlines are shown in figure \ref{fig:egv-xy}(a,b), is seen
to have a single row of recirculation cells along the edge. Figure
\ref{fig:egv-x}(a,b) shows that each subsequent mode has one more
row of cells, which are separated by the zero crossings of $\hat{v}.$
The perturbation of the normal magnetic field in the quasi-stationary
limit, which is shown in figure \ref{fig:egv-x}(e), has a characteristic
logarithmic singularity similar to that of the base field (\ref{eq:B0}).
This singularity is presumably behind the slow convergence noted above.
As seen in figure \ref{fig:egv-x}(f), this singularity vanishes in
the ideally conducting limit. For a negligible magnetic diffusion,
equation (\ref{eq:hhn}) at the edge $(x\rightarrow0)$ reduces to
$\gamma\Sl^{2}\hat{B}(0)=\hat{v}'(0)/2\pi,$ which implies a finite
normal magnetic field $\hat{B}(0)$ proportional to the longitudinal
velocity $\hat{u}(0)=i\hat{v}'(0)$ along the edge. But this, in turn,
means a singularity in the longitudinal current density, which is
seen in figure \ref{fig:egv-x}(d) to increase as $\sim x^{-1/2}$
towards the edge. Such a singular current distribution is typical
at the edges of perfectly conducting sheets, where it generates a
purely tangential magnetic field \citep{Priede2006,Priede2011a}.
The current singularity is obviously smoothed out by finite magnetic
diffusion, which thus produces a normal magnetic field through the
sheet with a logarithmic singularity at the edge. The associated spatial
amplitude distribution of the scalar magnetic potential $\hat{\Psi}(x,z)$
may be seen in figure \ref{fig:psi-xz} to have a symmetric discontinuity
over the sheet with a continuous normal derivative which defines the
normal magnetic field through the sheet. In the ideally conducting
limit shown in figure \ref{fig:psi-xz}(b), the perturbation is entirely
due to the advection of the magnetic field and, thus, more localized
at the edge. In the quasi-stationary limit shown in figure \ref{fig:psi-xz}(a),
the perturbation is spread out over the sheet by the magnetic diffusion.

As may be seen in figure \ref{fig:egv-x}, the increased current density
at the edge is produced by the flow towards the edge. It means that
the flow effectively compresses the current lines where it is directed
towards the edge, while the opposite is the case where the flow is
directed away from the edge. As this happens between the vortices,
the associated pattern of current lines in figure \ref{fig:egv-xy}(d,e)
is shifted by a quarter wave length relative to the streamline pattern
in figure \ref{fig:egv-xy}(a,d).

The compression of the current lines occurs as follows. According
to equation (\ref{eq:h}), a flow in a non-uniform magnetic field
induces electric current in order to conserve the magnetic flux carried
by the liquid. Since the magnetic field drops off with the distance
from the edge, the flow towards the edge locally reduces the magnetic
flux density, which may be seen in figure \ref{fig:egv-x}(f) for
the perfectly conducting limit. This, in turn, implies an increase
in the current density at the edge. The current density at the edge
has to raise in order to compensate the magnetic field generated by
the rest of the sheet. The increased local current density interacting
with the base magnetic field enhances the pinch force. This effect
outweighs the advection of weaker magnetic field towards the edge.
As a result, the initial flow perturbation towards the edge is enhanced,
thus giving rise to instability in ideally conducting liquid.

Additional amplification of the instability results from the magnetic
diffusion, which, as discussed above, smooths out the current singularity,
so producing a logarithmic singularity in the normal magnetic field
at the edge. As seen in figure \ref{fig:egv-x}(e), this perturbation
is positive and, thus, enhancing the base field at the edge. Therefore,
the growth rate of instability increases as the magnetic diffusion
becomes stronger and the Lundquist number respectively smaller. As
seen in figure \ref{fig:grate}(b), the growth rate saturates at $\Sl\lesssim10,$
which means that the instability becomes dominated by the magnetic
diffusion. In this range of $\Sl,$ the instability grows on the magnetic
response time scale (\ref{eq:tau-m}), which by a factor of $\Sl^{-1}$
exceeds the respective Alfvén time scale (\ref{eq:t-alfv}). The latter
is seen to apply only to $\Sl\gtrsim10.$

\section{\label{sec:Sum}Summary and conclusions}

In this study, we considered a pinch-type instability in a semi-infinite
planar sheet of an inviscid incompressible liquid with a straight
rigid edge carrying a uniform tangential electric current. The electromagnetic
pinch force resulting from the interaction of the electric current
with its own magnetic in this model is balanced by the pressure gradient
in quiescent liquid. This equilibrium state was found to be inherently
unstable with respect to the internal flow perturbation caused by
a row of counter-rotating vortices along the edge. The vortices compress
the electric current lines where the flow is directed towards the
edge and disperse them in the opposite case. The compression and dispersion
of the current lines occurs as a weaker magnetic field is advected
by the flow towards the edge and a stronger field is carried away.
The interaction of the current perturbation with the base magnetic
field produces a pinch force that amplifies initial flow perturbation.
In an ideally conducting liquid, this part of the pinch force dominates
over the opposing one which results from the interaction of the magnetic
flux perturbation with the base current. 

The role of the latter changes in the resistive limit, where it becomes
destabilizing and dominant. The difference in the instability mechanism
is due to the current and the magnetic field distributions at the
edge. In an ideally conducting liquid, perturbation of the magnetic
field remains finite at the edge, whereas the associated current density
increases as $\sim x^{-1/2}$ becoming unbounded at the edge. In a
resistive liquid, this current singularity is smoothed out by the
magnetic diffusion, which thus produces a logarithmic singularity
in the magnetic field. The latter enhances the base magnetic field
at the edge. This gives rise to a destabilizing perturbation of the
pinch force which dominates in the highly resistive liquid over the
interaction of the current perturbation with the base magnetic field. 

Different instability mechanisms lead to disparate instability development
times in the highly resistive and well conducting fluids. In the former,
the instability develops on the magnetic response time scale, which
depends on the conductivity. In the latter, the instability develops
on the much shorter Alfvén time scale. This is because the perturbation
of the magnetic field in poorly conducting fluid depends on the product
of conductivity and velocity, whereas in well conducting fluid the
magnetic field is effectively frozen in and, thus, its perturbation
depends directly on the displacement. Therefore these two instability
development times are generic and, thus, expected to hold also in
other geometries with fixed boundaries regardless of singularities
in the magnetic field or the electric current distributions. 

It follows from our model that the relevant electromagnetic parameter
of the problem is the linear current density. In the eutectic alloy
of GaInSn with $\rho=\unit[6.4\times10^{3}]{kg/m^{3}}$ and $\sigma=\unit[3.3\times10^{6}]{S/m}$
\citet{Weber2013} carrying linear current density of $J_{0}=\unit[1]{kA/cm},$
which is comparable to the respective quantity for the $m=1$ instability
mode in the cylinder with the radius of $\unit[5]{cm}$ and the critical
current of $\unit[2.9]{kA}$ \citep{Ruediger2011,Seilmayer2012},
characteristic instability development time is $\tau_{m}/\gamma_{1}\approx30\rho/(\sigma\mu_{0}^{2}J_{0}^{2})\approx\unit[4]{s},$
where $\gamma_{1}\approx0.0335$ is the dimensionless growth rate
of the most unstable resistive mode. Note that in a Li-Te liquid metal
battery, where the areal current density can reach up to $j_{0}=\unit[7]{A/cm^{2}}$
\citep{Kim2013}, the linear current density $J_{0}=\unit[1]{kA/cm}$
corresponds to a layer of thickness $\delta\sim\unit[1]{m.}$ The
respective instability development time in Li $(\sigma=\unit[3.3\times10^{6}]{S/m},$
$\rho=\unit[0.5\times10^{3}]{kg/m^{3}}$ \citep{Mueller2001} is then
by an order of magnitude shorter than that in GaInSn. Comparing the
instability development time in GaInSn with the viscous damping time
$\tau_{\nu}=\frac{\delta^{2}}{12\nu}$ due to the kinematic viscosity
$\nu$ in the respective Hele-Shaw flow \citep{Batchelor1967}, we
can estimate critical Hartmann number \global\long\def\Ha{\mathrm{Ha}}
$\Ha_{c}=\mu_{0}J_{0}\delta\sqrt{\sigma/\rho\nu}=\sqrt{12/\gamma_{1}}\approx20.$
This value is comparable with $\Ha{}_{c}\approx25$ for the non-axisymmetric
($m=1)$ instability mode in the cylindrical geometry \citep{Ruediger2011},
and it is also not far from $\Ha_{c}\approx40$ for the axisymmetric
mode in the annular geometry \citet{Priede2015}. For $\Ha\gtrsim10,$
a more adequate estimate may be provided by the so-called Hartmann
damping time $\tau_{H}=\frac{\delta^{2}}{2\nu}\Ha^{-1}$ \citep{Sommeria1982},
which is based on the Hartmann rather than Poiseuille velocity profile,
as in the Hele-Show flow, and yields $\Ha_{c}=2/\gamma_{1}\approx60.$
For liquid metals, which are relatively poor conductors characterized
by a low magnetic Prandtl number $\Pm=\mu_{0}\sigma\nu\sim10^{-5}-10^{-6},$
all these critical Hartmann numbers $\sim10$ correspond to the Lundquist
number $\Sl=\Ha\Pm^{1/2}\lesssim0.1,$ which means a highly resistive
mode of instability. 

In conclusion, note that in a viscous fluid, the threshold of instability,
which is defined by $\gamma=0,$ depends only the Hartmann number
\citep{Ruediger2011}. This is because the magnetic Prandtl number
vanishes from equations (\ref{eq:vhn}) and (\ref{eq:hhn}) together
with the Lundquist number similarly to the thermal Prandtl number
in the case of Rayleigh-Bénard instability \citep{Chandrasekhar1961}.
But it is important to stress that the actual mechanism of the instability
and, in particular, its growth rate depend on the Lundquist number.
Also non-linear evolution of the instability depends not on only on
the Hartmann number but also on the Lundquist (magnetic Prandtl) number
as recently demonstrated by \citet{Herreman2015}. 

\begin{acknowledgements}
This work was supported by Helmholtz Association of German Research Centres (HGF) in the framework of the LIMTECH Alliance through an agreement between Coventry University and Helmholz-Zentrum Dresden-Rossendorf.
\end{acknowledgements}

\bibliographystyle{jfm}
\bibliography{edge}

\end{document}